\documentclass{owrart}
\usepackage{amsmath}
\usepackage{amssymb}
\usepackage{graphicx} 
\newcommand{\aset}{{\mathbb{S}}}
\newcommand{\card}[1]{{| #1 |}}
\newcommand{\continuous}{\mathcal{C}}
\newcommand{\CP}{\hat{C}}
\newcommand{\CX}{C}
\newcommand{\de}{\mathrm{d}}
\newcommand{\ES}{\mathcal{E}}
\newcommand{\flowtot}[2]{{{{{\mathcal{F}}_{#1}^{#2}}}}}
\newcommand{\integers}{\mathbb{Z}}
\newcommand{\internal}{{\mathcal{I}}}

\newcommand{\LP}{\hat{\mathcal{L}}}

\newcommand{\logp}{{\log_{\scriptscriptstyle +}}}
\newcommand{\lra}[1]{\left\langle #1 \right\rangle}
\newcommand{\lrB}[1]{\left\{ #1 \right\}}
\newcommand{\lrp}[1]{\left( #1 \right)}

\newcommand{\bigglrp}[1]{\biggl( #1 \biggr)}

\providecommand{\lvert}{|}
\providecommand{\rvert}{|}
\newcommand{\lrv}[1]{\left\lvert #1 \right\rvert}
\newcommand{\slrv}[1]{\lvert #1 \rvert}

\newcommand{\miw}{w}
\newcommand{\naturals}{\mathbb{N}}

\newcommand{\NE}{\mathtt{N}}

\newcommand{\NL}{{\mathtt{L}}}

\newcommand{\NR}{{\mathtt{R}}}
\newcommand{\pa}{\partial}
\newcommand{\polyP}{\mathcal{P}}

\newcommand{\reals}{\mathbb{R}}
\newcommand{\rmi}{\mathrm{i}}

\newcommand{\ssty}{\scriptstyle}
\newcommand{\sssty}{\scriptscriptstyle}
\newcommand{\sX}{{\varphi}}
\newcommand{\Sint}{S^{\mathrm{int}}}

\newcommand{\tree}{{\tau}}
\newcommand{\wrho}{{\rho}}
\newcommand{\wtheta}{{\theta}}

\newcommand{\Wtrees}{{\mathcal{T}}}

\begin{document}
\input{obw_titlepage.sty}

\begin{talk}[Christoph Kopper,\cite{rg_guida_kopper}]{Riccardo Guida}
{All--order uniform momentum bounds for the massless
$\phi^4$ theory in four dimensional Euclidean space}
{Guida, Riccardo}

\noindent
A panoramic overview is given,
of a theorem \cite{rg_guida_kopper}
establishing physical and uniform bounds on the
Fourier--transformed Schwinger functions of a massless
$\phi^4$ theory in four Euclidean dimensions, at any
loop order in perturbation theory.

The first step to set up the perturbative framework
is to specify a free quantum theory
describing a massless scalar field
by fixing a centered Gaussian measure on
$\mathcal{S}'(\reals^4)$,
$\mu_{\ssty\hbar\CX^{\sssty\Lambda,\Lambda_0}_{R}}$,
whose covariance
$\hbar\CX^{\sssty\Lambda,\Lambda_0}_{R}(x,y)
:=\hbar\chi_R(x)\,\chi_R(y)
\,\CX^{\sssty\Lambda,\Lambda_0}(x-y)
$ is
assumed to be a distribution in $\mathcal{S}'(\reals^8)$
acting as a positive bilinear form on test functions.
$\hbar>0$ denotes
the variable of the formal perturbative series.
The short--distance behavior (smoothness)
of $\CX^{\sssty\Lambda,\Lambda_0}(x)$
as a function is controlled by
$\Lambda_0>0$ (known as ultra--violet, UV, cutoff),
while the long--distance regularity is controlled by
$0<\Lambda\le\Lambda_0$ (infra--red, IR, cutoff).
$\CX^{\sssty\Lambda_0,\Lambda_0}$ vanishes.
When $\Lambda_0$ tends to infinity
and $\Lambda$ tends to zero,
$\CX^{\sssty\Lambda,\Lambda_0}(x)$
approaches the standard free propagator
$\lra{x\lrv{\partial^{-2}}0}$.
For any $R>0$, the non--negative function
$\chi_R\in\continuous^{\infty}_c\lrp{\reals^4}$
satisfies the ``finite--volume'' constraint
$\chi_R(x)=1$ for any $|x|\le R$.

For any $\NE\in \naturals$, and any $\NL\in\naturals_0$
the Schwinger functions in momentum space are defined by
\begin{align}\label{rg.SF}
&\LP_{\NE,\NL}^{\sssty\Lambda,\Lambda_0}(p_{[\NE-1]})
:=\!\lim_{R\rightarrow \infty}
\biggl\lgroup
\bigglrp{\frac{1}{\NL!}\frac{\pa^\NL}{\pa\hbar^\NL}}_{\hbar=0}
\bigglrp{
\frac{\delta}{\delta\sX(0)}
\;\prod_{e=1}^{\NE-1}
\int\de^4 x_e\; e^{- \rmi x_e p_e}
\frac{\delta}{\delta\sX(x_e)}}_{\sX=0}
\\\nonumber
&\qquad
\bigglrp{
-\hbar\log\lrp{
\int\de\mu_{\hbar\CX^{\sssty\Lambda,\Lambda_0}_{R}}(\phi)
e^{-\frac{1}{\hbar}\Sint(\phi+\sX)}/
\int\de\mu_{\hbar\CX^{\sssty\Lambda,\Lambda_0}_{R}}(\phi)
e^{-\frac{1}{\hbar}\Sint(\phi)}}
}
\biggr\rgroup,
\end{align}
where:
$[a]:=[1:b]$,
$[a:b]:=\lrB{n\in\integers|a\le n\le b}$,
and
$p_{[n]}:=(p_1,\cdots,p_n)$.
In~\eqref{rg.SF}, the interaction action $\Sint(\sX)$ 
is defined by
\begin{align}\label{rg.Sint}
\Sint(\sX):=\int \de^4x\lrp{
A(\hbar)\,\frac{(\pa\sX(x))^2}{2}
+B_2(\hbar)\frac{\sX(x)^2}{2}
+B_4(\hbar)\frac{\sX(x)^4}{4!}
},
\end{align}
where $A,B_2,B_4$ are formal series in $\hbar$,
whose coefficients are fixed order by order by
appropriate renormalization conditions, in such a way 
that the ``UV+IR limit''
$
\lim_{\Lambda_0\rightarrow \infty}
\lim_{\Lambda\rightarrow 0^+}
\LP_{\NE,\NL}^{\sssty\Lambda,\Lambda_0}
$
exists in $\mathcal{S}'(\reals^{4(\NE-1)})$ for all $\NE,\NL$.
In particular, it turns out for a massless theory that
$A,B_2$ are of order $O(\hbar)$,
while $B_4=g_0+O(\hbar)$.
From~\eqref{rg.SF} and~\eqref{rg.Sint} it
follows that
$\LP_{2,0}^{\sssty\Lambda,\Lambda_0}$
and all $\LP_{\NE,\NL}^{\sssty\Lambda,\Lambda_0}$
with odd $\NE$ vanish.

The UV+IR limit of $\LP_{\NE,\NL}^{\sssty\Lambda,\Lambda_0}$
is a regular function only at non--exceptional momenta,
see e.g. \cite{rg_keller_kopper}.
(A collection of four vectors
$p_{[\NE-1]}$ is said \textit{exceptional} iff it exists
a non--empty $\aset\subseteq[\NE-1]$ such that
$\sum_{e\in\aset}p_e=0$.)

Any Schwinger function
$\LP_{\NE,\NL}^{\sssty\Lambda,\Lambda_0}$
defined in~\eqref{rg.SF}
can be computed from the standard weighted sum
of all Feynman amplitudes proportional to $\hbar^{\NL}$,
obtained via Feynman rules from an appropriate set of
connected amputated graphs with $\NE$ external lines.
Each such set includes all graphs
with vertices of coordination number $4$ and loop number
$\NL$.
The word ``amputated''
means that  Feynman rules do not
associate any factor to the external lines. 

Schwinger functions satisfy the 
``Polchinski'' renormalization group (RG) flow equations,
\cite{rg_polchinski}
(see \cite{rg_mueller} for an introduction),
which in their perturbative form read:
\begin{flalign}
\label{rg.eqrgflowsym}
\pa_\Lambda\LP_{\NE,\NL}^{\sssty\Lambda,\Lambda_0}
\lrp{p_{[\NE-1]}}
=\flowtot{\NE,\NL,\miw}{\sssty\Lambda,\Lambda_0}
:=\Bigl\lgroup
\frac{1}{2}\int\frac{\de^4 \ell}{(2\pi)^4}\,
\pa_\Lambda \CP^{\sssty\Lambda,\Lambda_0}\lrp{\ell}\,
\LP_{\NE+2,\NL-1}^{\sssty\Lambda,\Lambda_0}
\lrp{p_{[\NE-1]},-\ell,\ell}
\\
-\frac{1}{2}
\sum_{
  \begin{subarray}{c}
     {\ES'}\uplus\,{\ES''}=\,{[0:\NE-1]}
  \\
     {\NL'}+{\NL''}=\,\NL
  \end{subarray}}
  \!\!\!\!\!\!
\pa_\Lambda \CP^{\sssty\Lambda,\Lambda_0}
  \lrp{{\ssty\sum_{e\in\ES'}p_e}}\;
\LP_{\NE',\NL'}^{\sssty\Lambda,\Lambda_0}
  \lrp{p_{{\ES'}}}\;
\LP_{\NE'',{\NL''}}^{\sssty\Lambda,\Lambda_0}
 \lrp{p_{{\ES''}}}
 \Bigr\rgroup
,
\nonumber
\end{flalign}
where $\NE':=\card{{\ES'}}+1$, $\NE'':=\card{{\ES''}}+1$,
$p_0:=-\sum_{e\in[\NE-1]}p_e$,
and the sum on the r.h.s.
of~\eqref{rg.eqrgflowsym}
runs over all disjoint
(possibly empty) sets $\ES',\ES''$ whose union gives
$[0:\NE-1]$,
as well as over all non--negative integers
$\NL',\NL''$ whose sum gives $\NL$.

When the field has a mass $m>0$, it is not 
difficult to use the RG equations
to bound Schwinger functions in momentum space
(see e.g. \cite{rg_mueller}). 
Such bounds are simple but clearly unphysical
because they depend polynomially on external momenta;
moreover, they diverge when the mass vanishes
and the IR limit is taken.
More physical bounds have been proved in the massive case,
\cite{rg_kopper_meunier}.

The goal of the ``existence and boundedness theorem''
in \cite{rg_guida_kopper}
is to extend the ideas in 
\cite{rg_kopper_meunier} to obtain physical, uniform bounds
for the massless case.
The theorem assumes
that the Fourier--transformed covariance
$\CP^{\sssty\Lambda,\Lambda_0}(p)$
is
$O(4)$ invariant,
smooth in some sense, and such that 
$\Lambda^3\pa_\Lambda\CP^{\sssty\Lambda,\Lambda_0}(p)$
and
$\Lambda_0^2\Lambda^2
\pa_\Lambda\pa_{\Lambda_0}\CP^{\sssty\Lambda,\Lambda_0}(p)$
(together will all necessary derivatives w.r.t. $p$) 
are exponentially decreasing when
$|p|/\Lambda\rightarrow \infty$.
The main result  of the theorem is that
for any $\NE,\NL$ and any
multi--index $\miw\in \naturals_0^{4(\NE-1)}$,
there exist
a polynomial $\polyP_{\NL}$
of degree $\le\NL$ and with non--negative coefficients,
as well as
a set of weighted trees $\Wtrees_{\NE,2\NL,\miw}$,
such that (when e.g. $\NE\ge 4$)
\begin{align}\label{rg.zeromass}
\lrv{
\pa^\miw_p\LP_{\NE\ge4,\NL}^{\sssty\Lambda,\Lambda_0}
\lrp{p_{[\NE-1]}} }
&\le
\polyP_{\NL}
\bigglrp{
  \logp\bigglrp{
    \frac{|p_{[\NE-1]}|_\mu}
         {\kappa}}
 ,\logp{\frac{\Lambda}{\mu}}
}\sum_{T\in\Wtrees_{\NE,2\NL,\miw}
}
\prod_{i\in\internal\lrp{T}} 
|{k_i}|_\Lambda^{-\wtheta(i)}
\end{align}
for any $\Lambda_0>0$, $0<\Lambda\le\Lambda_0$
and $p_{[\NE-1]}\in \reals^{4(\NE-1)}$.
In \eqref{rg.zeromass},
$\mu>0$ is the renormalization scale;
$|p_{[\NE-1]}|:=\sup_e |p_e|$;
$\lrv{p}_\Lambda:=\sup(\Lambda,\lrv{p})$;
$\logp{x}:=\log\sup(1,x)$.
$\kappa:=\sup(\Lambda,\inf(\eta(p_{[\NE-1]}),\mu))>0$
is defined in terms of a ``dynamical IR cutoff''
$\eta(p_{[\NE-1]}):=
\inf_{\emptyset\neq\aset\subseteq[\NE-1]}
|\sum_{e\in\aset}p_e|$
(positive for non--exceptional momenta).
$\internal\lrp{T}$ is the set of internal lines of the
weighted tree $T$;
$k_i$ is the momentum flowing through the internal line $i$,
and $\wtheta(i)>0$ is the total weight
associated to $i$.

The sets
$\Wtrees_{\NE,\NR,\miw}$ ($\NR\in \naturals_0$)
satisfy two properties;
\textit{nestedness:}
$\Wtrees_{\NE,\NR,\miw}\subseteq\Wtrees_{\NE,\NR+1,\miw}$;
\textit{saturation:}
$\Wtrees_{\NE,\NR,\miw}=\Wtrees_{\NE,3\NE-2,\miw}$
for any $\NR\ge 3\NE-2$.
The set $\Wtrees_{\NE,\NR,\miw=0}$
(corresponding to the absence of derivatives
w.r.t. external momenta)
is defined
as the set of all $T=(\tree,\wrho)$ in which $\tree$
is a tree and 
$\wrho:\internal\lrp{T}\rightarrow\lrB{1,2}$
is a line weight, such that:
\textit{a)} $\tree$ has
$\NE$ external lines and vertices of coordination number
in $\lrB{3,4}$;
\textit{b)} the number of vertices with coordination $3$
is $\le\NR$;
\textit{c)} $\sum_{i\in\internal\lrp{T}}\wrho(i)=\NE-4$;
\textit{d)} there is a bijection among the vertices of
coordination number $3$ and the internal lines with
$\wrho=1$. In the case  $\miw=0$ one has $\wtheta(i)=\wrho(i)$.

As an example, for any $\NL>0$ the set  
$\Wtrees_{\NE=6,\NR=2\NL,\miw=0}$ contains only the trees
\begin{align*}
&
\begin{minipage}[c]{0.23\textwidth}
\includegraphics[width=\textwidth]{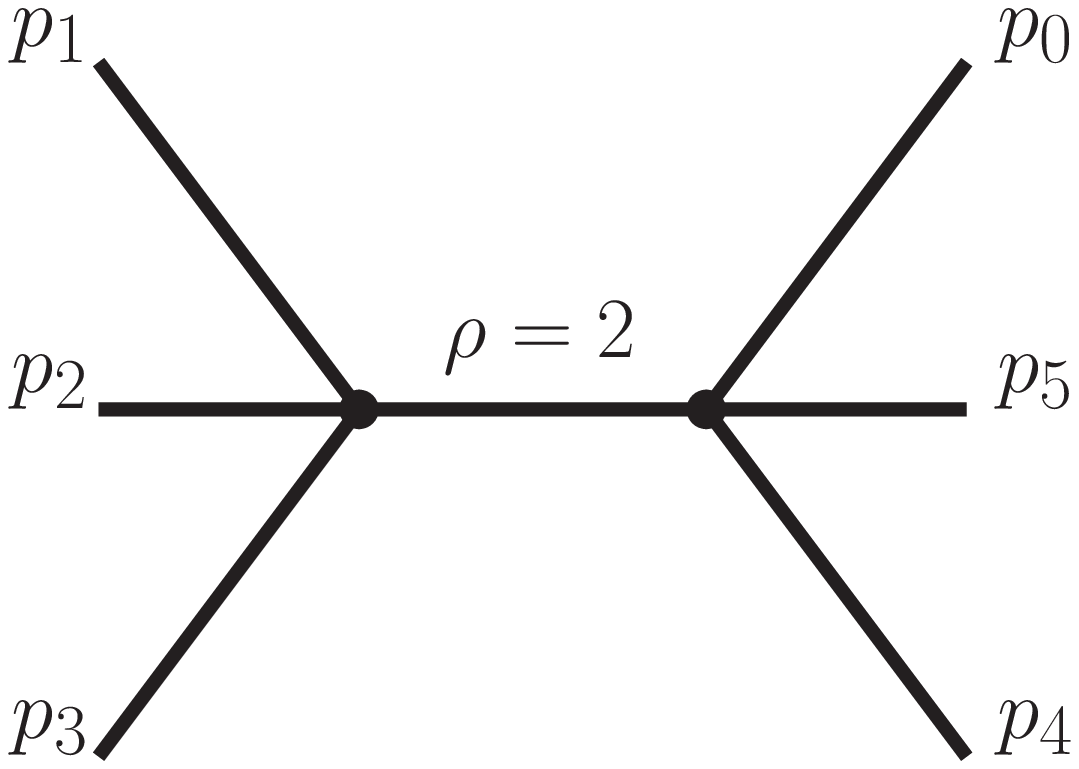}
\end{minipage},\quad
\begin{minipage}[c]{0.29\textwidth}
\includegraphics[width=\textwidth]{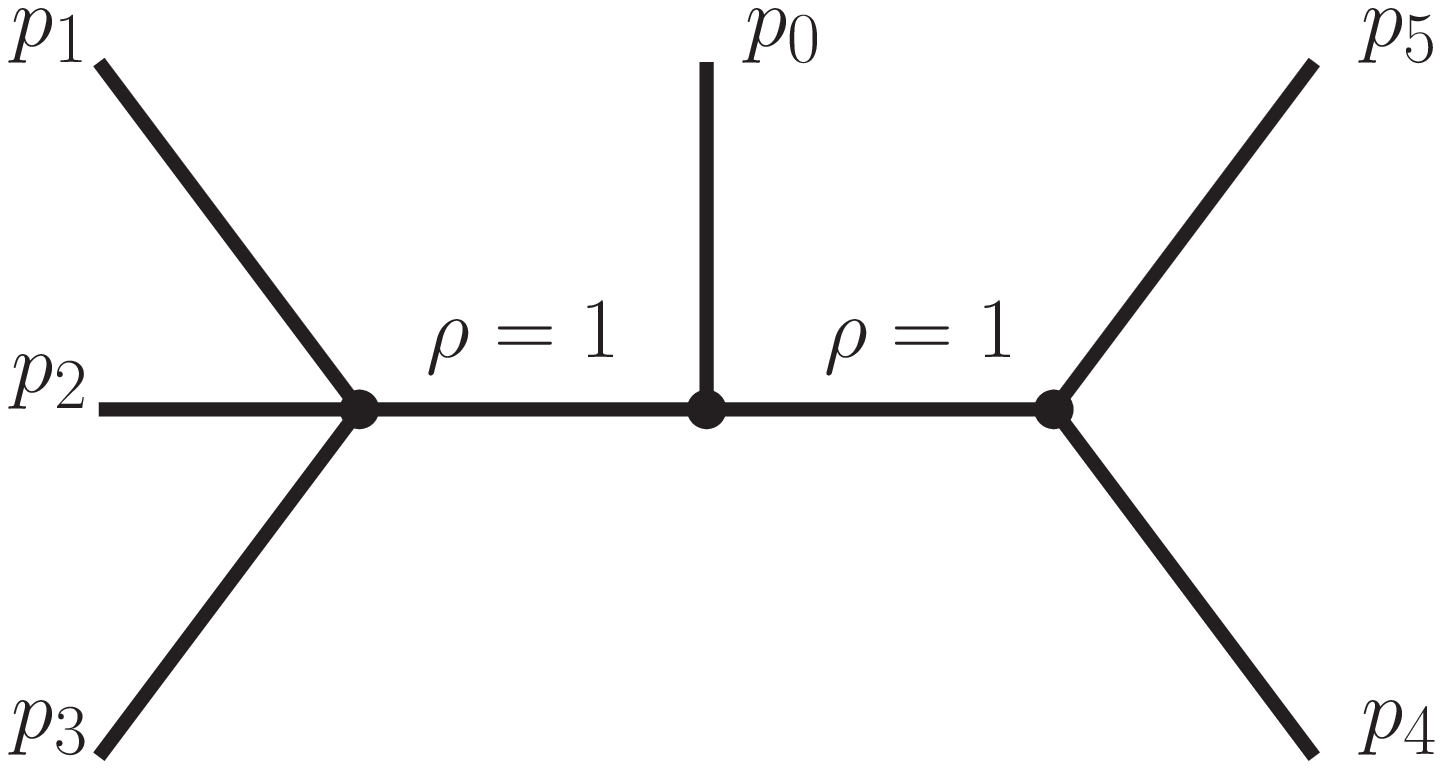}
\end{minipage},\quad
\begin{minipage}[c]{0.29\textwidth}
\includegraphics[width=\textwidth]{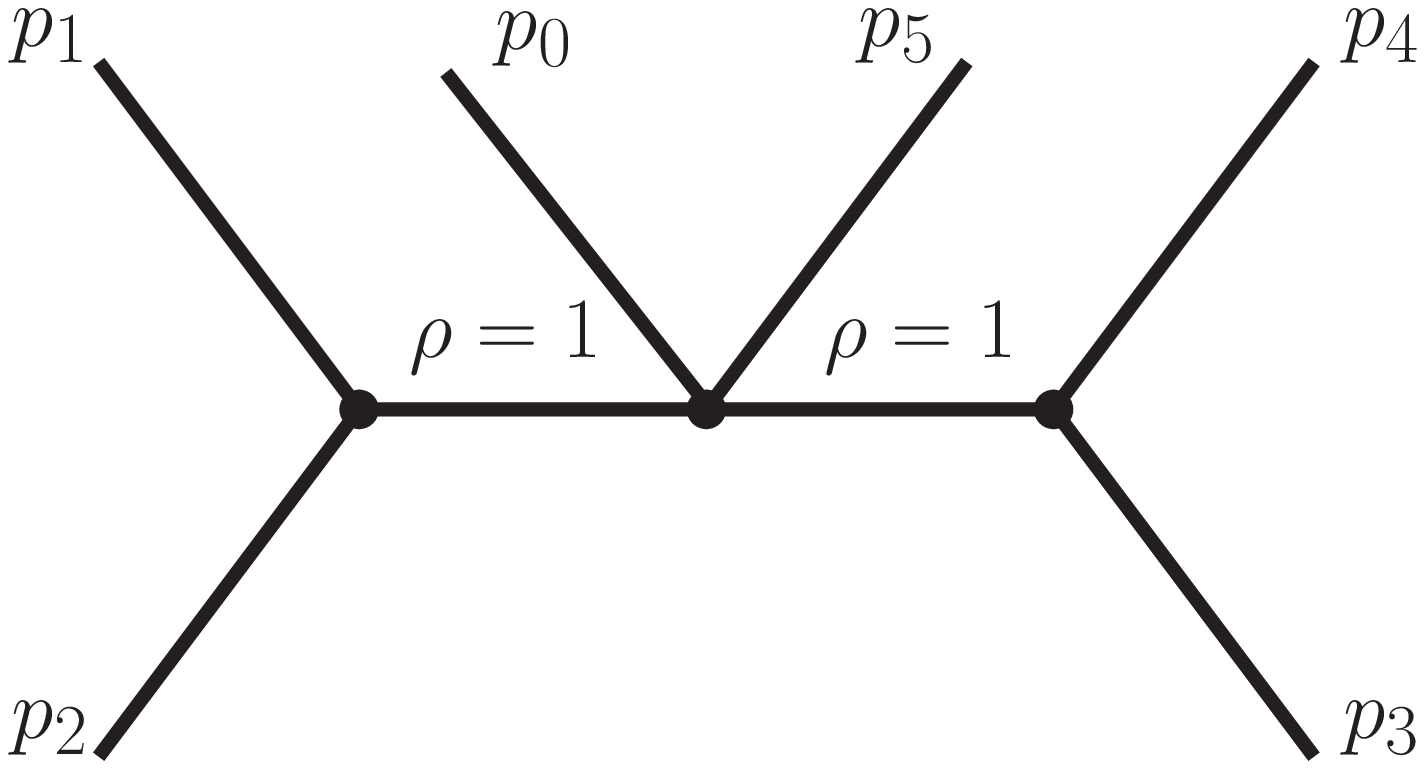}
\end{minipage},
\end{align*}
and the trees derived from them
by non--trivial
permutations of the external momenta $p_{[0:5]}$.
(Other trees with $\NE$ external lines and 
vertices of coordination numbers $3,4$ exist
but do not satisfy to the defining conditions.)
Correspondingly, in this case
the bound~\eqref{rg.zeromass} reads for any $\NL>0$ 
\begin{align*}
\slrv{
\LP_{6,\NL}^{\sssty\Lambda,\Lambda_0}({p_{[5]}}) }
&\le
\lrp{\scriptstyle
{\lrv{p_1+p_2+p_3}_\Lambda^{-2}}
+
{\lrv{p_1+p_2+p_3}_\Lambda^{-1}}
{\lrv{p_4+p_5}_\Lambda^{-1}}
+
{\lrv{p_1+p_2}_\Lambda^{-1}}
{\lrv{p_3+p_4}_\Lambda^{-1}}
+\text{perms.}
}\polyP_{\NL},
\end{align*}
where $\polyP_{\NL}$ has been introduced
in~\eqref{rg.zeromass}.

The proof of the theorem
is based on the recursive structure of the
perturbative RG equations \eqref{rg.eqrgflowsym}
(see e.g. \cite{rg_mueller}).
The main difficulty is to wisely deal with spurious
exceptional
momenta, in order to keep the bound finite in the IR
limit.

In the flow
$\flowtot{\NE,\NL,\miw}{\sssty\Lambda,\Lambda_0} $,
see \eqref{rg.eqrgflowsym},
the term quadratic in Schwinger functions
acts as a junction
of the weighted trees $T',T''$ in the bounds,
respectively, of 
$\LP_{\NE',{\NL'}}^{\sssty\Lambda,
\Lambda_0}$,
$\LP_{\NE'',{\NL''}}^{\sssty\Lambda,
\Lambda_0}$.
Now, the junction of two weighted trees
happens to be a weighted tree of the appropriate class
and the inductive bound for
$\LP_{\NE,\NL}^{\sssty\Lambda,\Lambda_0}$ is then reproduced.

The linear term in 
$\flowtot{\NE,\NL,\miw}{\sssty\Lambda,\Lambda_0} $
is more problematic, because it contains a loop integration
which tends to destroy the tree structure of the bounds.
The exponential fall--off in
$\ell/\Lambda$ of the covariance allows to prove
(\cite{rg_kopper_meunier},\cite{rg_guida_kopper})
bounds of the form
\begin{align}\label{rg.XII}
\int\!\de^4\ell
\;\slrv{
\pa_\Lambda \CP^{\sssty\Lambda,\Lambda_0}
     \lrp{\ell} }
\,\prod_{j=1}^{n}
{\lrv{\ell+k_j}_\Lambda^{-\wtheta_j}}
\le\,c\,\Lambda\prod_{j=1}^{n}
{\lrv{k_j}_\Lambda^{-\wtheta_j}},
\end{align}
which, roughly speaking,
amount to ``cut the loop'' and to
set $\ell=0$ by deleting two external lines for each tree.
This property makes
the linear part of the flow more ``tree friendly''.
The elimination of the unwanted $\Lambda$ factor
in \eqref{rg.XII} (using the bound 
$\Lambda\le\slrv{k_{j'}}_\Lambda$ for some $j'$),
and the integration over $\Lambda$
(to recover Schwinger functions from the flow)
are taken into account by eliminating the factors
$\slrv{k_{j'}}_\Lambda^{-1},\slrv{k_{j''}}_\Lambda^{-1}$
for each tree in the original bound of
$\LP_{\NE+2,\NL-1}^{\sssty\Lambda,\Lambda_0}$,
which amounts to consider a subtraction of two
units in the original weights: this procedure can be 
consistently implemented as a mapping among our classes of
weighted trees.

The logarithms in \eqref{rg.zeromass} originate
from the $\Lambda$ integration of the flow
for marginal and irrelevant Schwinger functions,
as well as from the integral interpolating
marginal Schwinger functions from the renormalization point
to a generic one.

\end{talk}

\end{document}